\documentclass{iopart}
\usepackage{graphicx}

\begin{document}

\title[Zitterbewegung in semiconductors]{Zitterbewegung (trembling motion) of electrons in narrow gap
semiconductors}

\author{W Zawadzki\dag\ and T M Rusin*}

\address{ \dag Institute of Physics, Polish Academy of Sciences, 02-668 Warsaw, Poland \\
          $^*$PTK Centertel Sp. z o.o., 01-230 Warsaw, Poland}
 \ead{zawad@ifpan.edu.pl}
\begin{abstract}
Theory of trembling motion [Zitterbewegung (ZB)] of charge carriers in various narrow-gap materials
is reviewed. Nearly free electrons in a periodic potential, InSb-type semiconductors, bilayer graphene,
monolayer graphene and carbon nanotubes  are considered. General features of ZB are emphasized.
It is shown that, when the charge carriers are prepared in the form of Gaussian wave packets,
the ZB has a transient character with the decay time of femtoseconds in graphene
and picoseconds in nanotubes. Zitterbewegung of electrons in graphene in the presence of an external
magnetic field is mentioned. A similarity of ZB in semiconductors to that of relativistic
electrons in a vacuum is stressed. Possible ways of observing the trembling motion in solids are mentioned.
\end{abstract}


\section{Introduction}
Zitterbewegung (the trembling motion) was theoretically devised by Schroedinger \cite{Schroedinger30}
after Dirac had proposed his equation describing free relativistic
electrons in a vacuum. Schroedinger showed that, due to a non-commutativity of the quantum
velocity with the Dirac Hamiltonian, relativistic electrons experience the Zitterbewegung (ZB) even in
absence of external fields. The frequency of ZB is about $\omega = 2m_0c^2/\hbar$ and its
amplitude is about the Compton wavelength $\lambda_c=\hbar/m_0c \approx 3.86\times 10^{-
3}$\AA. It was later understood that the phenomenon of ZB is due to an
interference of electron states with positive electron energies ($E>m_0c^2$) and those with negative
energies ($E<m_0c^2$).
In other words, the ZB results from the structure of the Dirac
Hamiltonian, which contains both positive and negative electron energies. It is a purely
quantum effect as it goes beyond Newton's first law. To our knowledge, the ZB for free electrons has
never been directly observed. However,  in the presence of the
Coulomb potential the ZB is manifested in appearance of the so called Darwin term.
It was pointed out that the Zitterbewegung also may
occur in non-relativistic two-band systems in solids \cite{Cannata90}.
Since the appearance of papers by Zawadzki \cite{Zawadzki05KP}
and Schliemann \etal \cite{Schliemann05}
the subject of ZB became popular and it was demonstrated  that this phenomenon should
occur in various situations in solids
\cite{Katsnelson06,Cserti06,Zawadzki06,Winkler07,Rusin07a,Rusin07b,Rusin07c,Zuelicke08,Schliemann08}.
It is the purpose of the present review to outline main
features of ZB in narrow gap semiconductors.

We begin by elementary considerations based on the Schroedinger equation. In the absence of external
fields the Hamiltonian is $\hat{H}=\hat{p}^2/2m$ and the velocity is
$\hat{v}_j=\partial \hat{H}/\partial \hat{p}_j=\hat{p}_j/m$.
The time derivative of velocity is easily calculated
to give $\dot{\hat{v}}_j=1/(i\hbar)[\hat{v}_j,\hat{H}]=0$. This means that $\hat{v}_j(t)=$ const, which
is equivalent to first Newton's law stating that in absence of external forces the velocity is constant.

The situation is different when a Hamiltonian $\hat{H}$ has a matrix form, like in the Dirac equation for
relativistic electrons in a vacuum or in case of two or more interacting energy bands in solids.
Then the quantum velocity $\hat{v}_j=\partial \hat{H}/\partial \hat{p}_j$, which is also a matrix, does
not commute with the Hamiltonian and the quantum acceleration $d\hat{v}_j/dt$  does not vanish even
in the absence of external fields. Below we consider such situations and investigate their
consequences for the electron motion.

\section{Theory and results}

\begin{figure}
\includegraphics[width=7cm,height=7cm]{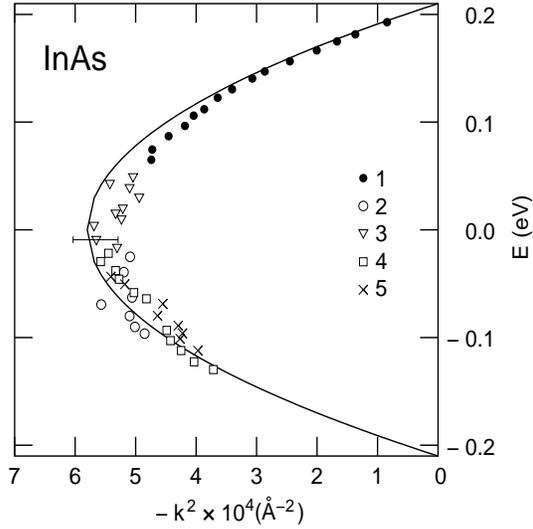}
\caption{ \label{Fig1}
Energy-wave vector dependence in the forbidden gap of InAs. Various symbols show the experimental data
of Parker and Mead (Ref. \cite{Parker68}). The solid line is a theoretical fit.
The determined parameters are $\lambda_Z=41.5$\AA\ and $u=1.33\times 10^8$ cm/s
(after Ref. \cite{Zawadzki05KP}).}
\end{figure}

We consider first the case of InSb-type narrow-gap semiconductors (NGS), see \cite{Zawadzki05KP}.
Their band structure is well described by
the model of $\Gamma_6$ (conduction), $\Gamma_8$ (light and heavy hole) and $\Gamma_7$ (split-off)
bands and it represents an $8 \times 8$ operator matrix. Assuming the spin-orbit
energy $\Delta \gg {\cal E}_g$, neglecting the free-electron terms and taking the momentum components
$\hat{p}_z\neq0$ and $\hat{p}_x=\hat{p}_y=0$, one obtains for the conduction and light hole bands the Hamiltonian
\begin{equation} \label{KP_H}
\hat{H} = u\hat{\alpha}_3\hat{p}_z + \frac{1}{2}{\cal E}_g\hat{\beta},
\end{equation}
where $\hat{\alpha}_3$ and $\hat{\beta}$ are the well known $4\times 4$ Dirac matrices
and $u=({\cal E}_g/2m_0^*)^{1/2}\approx 1\times 10^8$ cm/s is the maximum velocity.
Hamiltonian (\ref{KP_H}) has the form appearing in the Dirac equation.
The electron velocity is $\dot{\hat{z}}=(1/i\hbar)[\hat{z},\hat{H}]=u\hat{\alpha}_3$.
To determine $\hat{\alpha}_3(t)$ one calculates the commutator of $\hat{\alpha}_3$ with $\hat{H}$
and integrates the result with respect to time.
This gives $\dot{\hat{z}}(t)$, and $\hat{z}(t)$ is calculated integrating again. The result is
\begin{equation} \label{KP_z(t)}
\hat{z}(t) = \hat{z}(0) + \frac{u^2\hat{p}_z}{\hat{H}} + \frac{i\hbar u}{2\hat{H}} \hat{A}_0
 \left[  \exp\left( \frac{-2i\hat{H}t}{\hbar} \right)-1  \right],
\end{equation}
where $\hat{A}_0=\hat{\alpha}(0)-u\hat{p}_z/\hat{H}$. The first two terms represent the
classical motion. The third term describes time dependent oscillations with the frequency
$\omega_Z\approx {\cal E}_g/\hbar$. Since $\hat{A}_0\approx 1$, the amplitude of oscillations is
$\hbar u/\hat{H} \approx \hbar /(m_0^*u)=\lambda_Z$. Here
\begin{equation} \label{KP_lambda_z}
\lambda_Z = \frac{\hbar}{m_0^*u}
\end{equation}
is an important quantity analogous to the Compton wavelength $\lambda_c=\hbar/(m_0c)$ for relativistic
electrons in a vacuum. The oscillations analogous to those described in (\ref{KP_z(t)})
are called Zitterbewegung. The quantity $\lambda_Z$ can be measured directly.
The energy for electrons in NGS can be written in a 'semi-relativistic' form (see \cite{Zawadzki66})
$E= \pm \hbar u (\lambda_Z^{-2}+k^2)^{1/2}$. For $k^2>0$ this formula describes
the conduction and light-hole bands. For imaginary values of k there is $k^2<0$ and the above
formula describes the dispersion in the energy gap. This region is classically forbidden
but it can become accessible through quantum tunnelling. Figure \ref{Fig1} shows the data for
the dispersion in the gap of InAs, obtained from tunnelling experiments.
The fit gives $\lambda_Z\approx 41.5$\AA\  and
$u\approx 1.33\times 10^8$ cm/s, in good agreement with the estimation for InAs ($m_0^*\approx 0.024m_0$).
Similar data for GaAs give $\lambda_Z$ between 10 and 13 \AA\ \cite{Pfeffer90}, again in good agreement with
the theoretical predictions ($m_0^*\approx 0.066m_0$).

To demonstrate universality of the two-band situation we consider the well known case
of nearly free electrons in a solid in which the periodic lattice potential
$V(\bi{r})$ is treated as a perturbation (see \cite{Rusin07a}).
Near the Brillouin zone boundary the Hamiltonian
has, to a good approximation, a $2\times 2$ form
\begin{equation} \label{defhH}
 \hat{H} =  \left(\begin{array}{cc} \epsilon_{\bi{k}+\bi{q}}  & V_{\bi {q}} \\
                    V_{\bi{q}}^* & \epsilon_{\bi{k}} \\ \end{array}\right),
\end{equation}
where $V^*_{\bi{q}}= V_{-\bi{q}}$ are the Fourier coefficients in the expansion of $V(\bi{r})$,
and $\epsilon_{\bi k}= \hbar^2k_z^2/2m_0$ is the free electron energy. The $2\times 2$ quantum
velocity $\hat{v}_z$ can now be calculated and the acceleration $\dot{\hat{v}}_z$ is computed
in the standard way. Finally, one calculates the displacement matrix $\hat{z}_{ij}$.

Until now we treated the electrons as plane waves. However, Lock \cite{Lock79} in his important paper
observed that such a wave is not localized and it seems to be of a limited practicality to
speak of rapid oscillations in the average position of a wave of infinite extent.
Since the ZB is by its nature not a stationary state but a dynamical phenomenon, it is natural
to study it with the use of wave packets. These have become a practical instrument
when femtosecond pulse technology emerged (see Ref. \cite{Garraway95}). Thus, in a more realistic
picture the electrons are described by a wave packet
\begin{equation}\label{defPacket}
 \psi(z) = \frac{1}{\sqrt{2\pi}}\frac{d^{1/2}} {\pi^{1/4}}
           \int_{\infty}^{\infty}\exp\left(-\frac{1}{2}d^2(k_z-k_{z0})^2\right)
            \exp(i{k_zz}) dk_z   \left( \begin{array}{c}  1 \\  0\\\end{array} \right).  \ \ \ \ \ \ \
\end{equation}

\begin{figure}
\includegraphics[width=7cm,height=7cm]{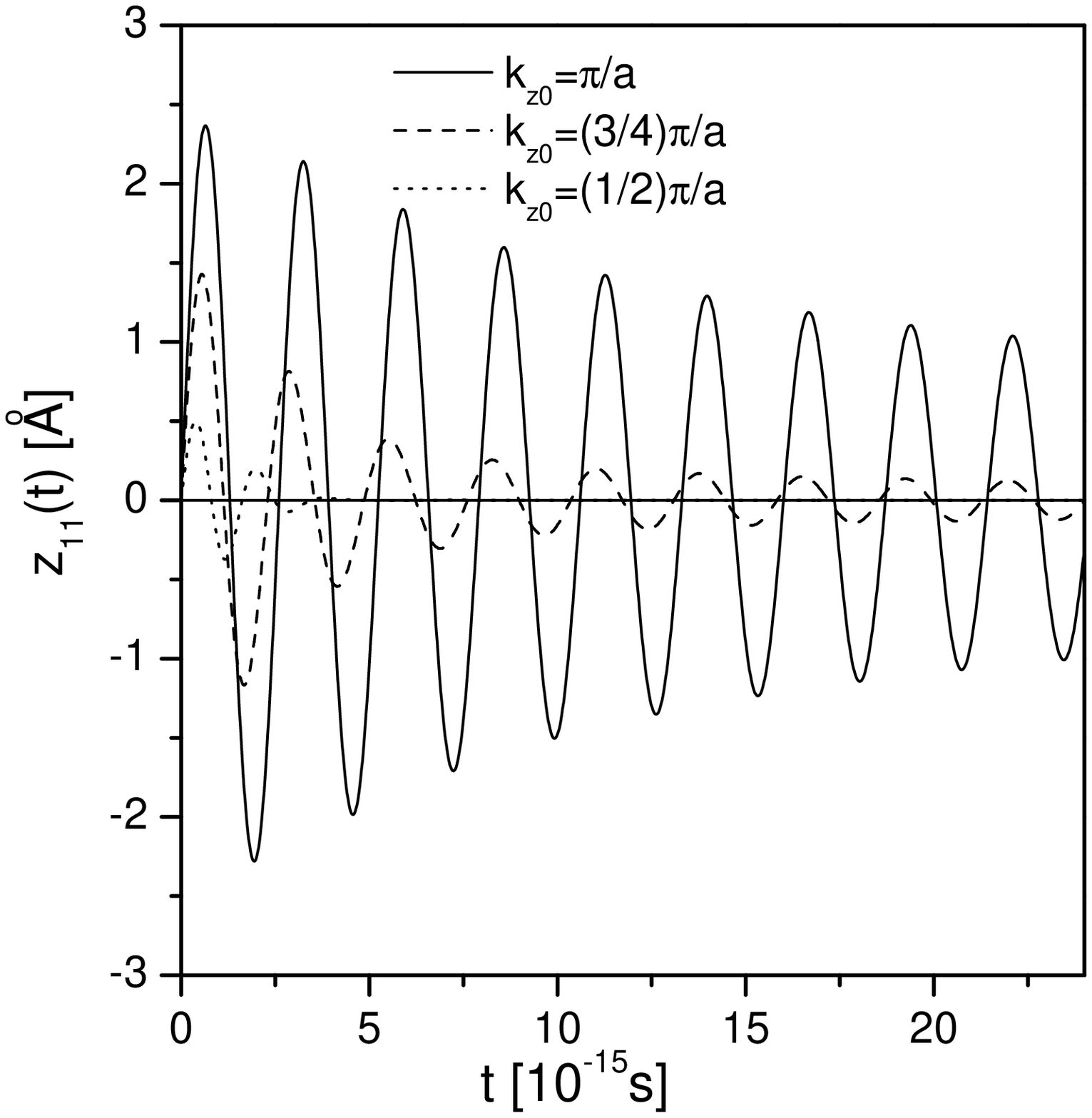}
\caption{\label{Fig2}  Transient Zitterbewegung oscillations of nearly-free electrons  {\it versus} time,
          calculated for a very narrow wave packet centered  at various $k_{z0}$ values.
          The band parameters correspond to GaAs.}
\end{figure}
In Figure \ref{Fig2} we show the ZB oscillations of $\hat{z}_{11}(t)$ averaged over wave packet (\ref{defPacket}).
The essential result is that, in agreement with Lock's general predictions, the ZB oscillations
of a wave packet have a {\it transient} character, i.e.
they disappear with time on a femtosecond scale.

\begin{figure}
\includegraphics[width=7cm,height=6cm]{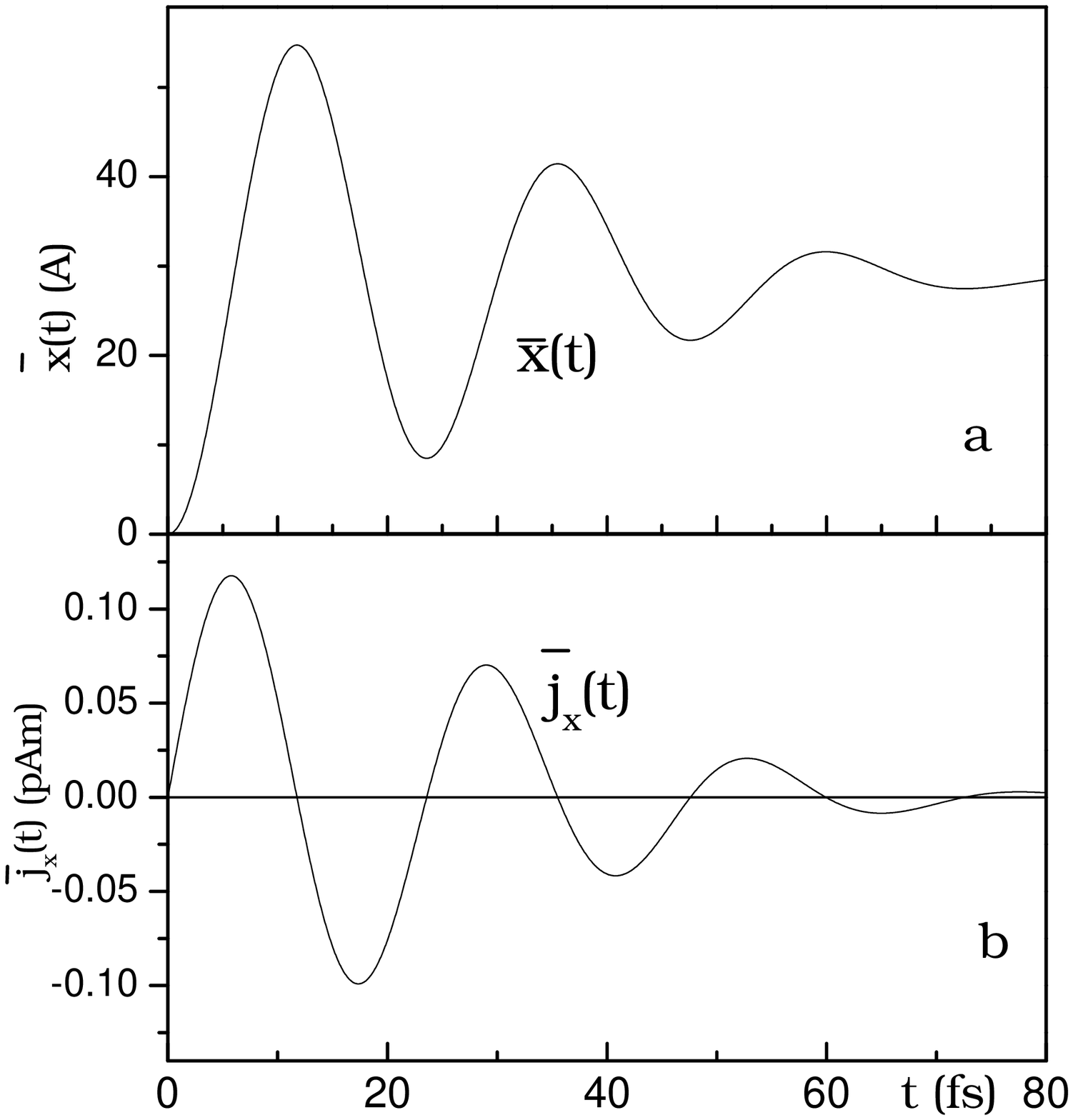}
\caption{ \label{Fig3} Zitterbewegung of a charge carrier in bilayer graphene {\it versus}
time, calculated for a Gaussian wave packet width $d=300$\AA\ and $k_{0y}=3.5\times 10^8$m$ ^{-1}$:
a) displacement, b) electric current.
The decay time is about $\Gamma_Z^{-1}=40$ fs
(after Ref. \cite{Rusin07b}).}
\end{figure}

Now we turn to interesting and intensively studied materials: bilayer and monolayer graphene and
carbon nanotubes (CNT), see \cite{Zawadzki06,Rusin07b,Novoselov04}.
Two-dimensional Hamiltonian for bilayer graphene is well approximated by \cite{McCann06PRL}
\begin{equation} \label{BG_H}
 \hat{H}_{B} = -\frac{1}{2m^*}\left(\begin{array}{cc}
     0 & (\hat{p}_x-i\hat{p}_y)^2 \\  (\hat{p}_x+i\hat{p}_y)^2 & 0 \\     \end{array}\right).
\end{equation}
The energy spectrum is ${\cal E}=\pm \hbar^2k^2/2m^*$, i.e.
there is no energy gap between the conduction and valence bands. The position operator in the
Heisenberg picture is a $2\times 2$ matrix
$\hat{x}(t) = \exp(i\hat{H}_Bt/\hbar)\hat{x}\exp(-i\hat{H}_Bt/\hbar).$
We calculate
\begin{equation} \label{BG_x_11}
x_{11}(t)= x(0)  + \frac{k_y}{k^2}\left[ 1 -\cos\left(\frac{\hbar k^2t}{m^*}\right)\right],
\end{equation}
where $k^2=k_x^2+k_y^2$.
The third term represents the Zitterbewegung with the frequency
$\hbar\omega_Z= 2\hbar^2k^2/2m^*$, corresponding to the energy difference
between the upper and lower energy branches for a given value of $k$.
In order to have the ZB in the $x$ direction one needs a non-vanishing
momentum $k_y$ in the $y$ direction. If the electron is represented by a two-dimensional
Gaussian wave packet similar to that given in Equation (\ref{defPacket}), the integrals of
interest can be calculated analytically. In Figure \ref{Fig3} we show observable physical quantities
related to ZB. It can be seen that they have a transient character.

Looking for the reason of transient character of ZB the electron wave packet is decomposed
into sub-packets corresponding to positive and negative electron energies. It turns out
that the 'positive' and 'negative' sub-packets move in the opposite
directions with the same velocity $v=\hbar k_{0y} t/2m^*$, where $\hbar k_{0y}$
is the initial value of momentum. The relative velocity
is $v^{rel}=\hbar k_{0y}t/m^*$. Each of these packets has the initial width $d$
and it spreads (slowly) in time. After the time $\Gamma_Z^{-1}=d/v^{rel}$
the distance between the two packets equals $d$, so the sub-packets cannot interfere and
the ZB amplitude diminishes. This reasoning gives the decay constant $\Gamma_Z=\hbar k_{0y}/m^*d$,
in agreement with the analytical results \cite{Rusin07b}.
Thus, the transient character of the ZB oscillations
in a collisionless sample is due to an increasing spacial separation of the sub-packets
corresponding to the positive and negative energy states.

\begin{figure}
\includegraphics[width=7cm,height=6cm]{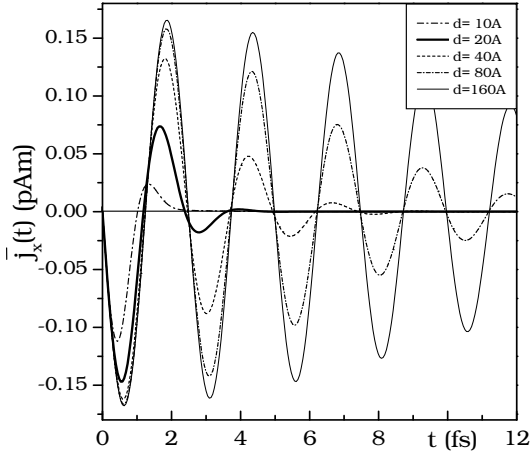}
\caption{ \label{Fig4} Oscillatory electric current in the $x$ direction
 caused by the ZB in monolayer graphene
{\it versus} time, calculated for a Gaussian wave packet with $k_{0y}=1.2\times 10^9$m$^{-1}$ and various
packet widths $d$ (after Ref. \cite{Rusin07b}).}
\end{figure}

Now, we turn to monolayer graphene.  The two-dimensional band Hamiltonian
describing its band structure is \cite{McClure56}
\begin{equation} \label{MG_H}
 \hat{H}_M = u\left(\begin{array}{cc}
     0 & \hat{p}_x-i\hat{p}_y \\  \hat{p}_x+i\hat{p}_y & 0 \\     \end{array}\right),
\end{equation}
where $u\approx 1\times 10^8$cm/s. The resulting energy dispersion is linear in momentum:
${\cal E}=\pm u\hbar k$, where $k=\sqrt{k_x^2+k_y^2}$.
The quantum velocity in the Schroedinger picture is
$\hat{v}_j=\partial \hat{H}_M/\partial \hat{p}_j$, it does not commute with the Hamiltonian.
In the Heisenberg picture we have
$\hat{\bi{v}}(t)=\exp(i\hat{H}_Mt/\hbar)\hat{\bi{v}}\exp(-i\hat{H}_Mt/\hbar)$.
Using Equation (\ref{MG_H}) we calculate
\begin{equation} \label{MG_v11}
 v_x^{(11)} = u \frac{k_y}{k}\sin(2ukt).
\end{equation}
The above equation describes the trembling motion with the frequency $\omega_Z=2uk$,
determined by the energy difference between the upper and lower energy branches for a given
value of $k$. As before, the ZB in the direction $x$ occurs only if there is a non-vanishing
momentum $\hbar k_y$. The results for the current
$\bar{j}_x=e\bar{v}_x$ are plotted in Figure \ref{Fig4} for $k_{0y}=1.2\times 10^9$m$^{-1}$
and different packet widths $d$. It is seen that the ZB frequency does not depend
on $d$ and is nearly equal to $\omega_Z$, as given above for the plane wave.
On the other hand, the amplitude of ZB does depend on $d$ and we deal with decay times
of the order of femtoseconds.

Finally, we consider monolayer graphene sheets rolled into single
semiconducting carbon nanotubes (CNT). The band Hamiltonian in
this case is similar to Equation (\ref{MG_H}) except that, because of
the periodic boundary conditions, the momentum $\hat{p}_x$ is quantized
and takes discrete values $\hbar k_x=\hbar k_{n\nu}$,
where $k_{n\nu}=(2\pi/L)(n-\nu/3)$, $n=0,\pm 1,\ldots$,
$\nu=\pm 1$, and $L$ is the length of circumference of CNT \cite{Ando93}.
As a result, the free electron motion can occur only in the direction
$y$, parallel to the tube axis. The geometry of CNT has two
important consequences. First, for $\nu=\pm 1$ there {\it always}
exists a non-vanishing value of the quantized momentum $\hbar k_{n\nu}$.
Second, for each value of $k_{n\nu}$ there exists $k_{-n,-\nu}=-k_{n\nu}$ resulting in the
same subband energy ${\cal E}=\pm E$, where $ E=\hbar u\sqrt{k_{n\nu}^2+k_y^2}$.

\begin{figure}
\includegraphics[width=7cm,height=7cm]{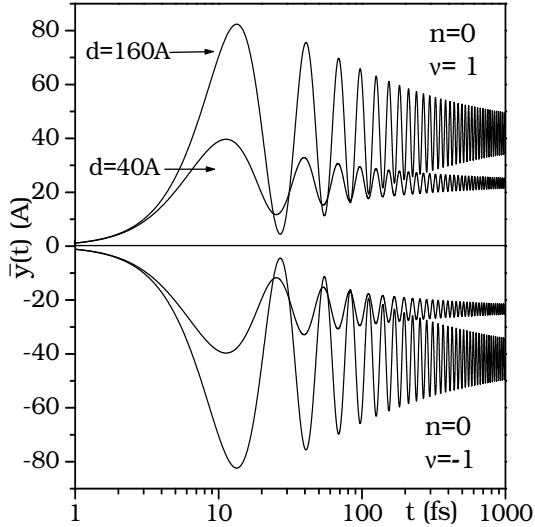}
\caption{ \label{Fig5}  Zitterbewegung of charge carriers in the ground
subband of a single carbon nanotube of $L=200$ \AA\ {\it versus} time (logarithmic scale),
calculated for Gaussian wave packets of two different
widths $d$ and $k_{0y}=0$. After the ZB disappears a constant shift remains. The two carriers
are described by different quantum numbers $\nu$ (after Ref. \cite{Rusin07b}).}
\end{figure}

The time dependent velocity $\hat{v}_y(t)$ and the displacement $\hat{y}(t)$ can be
calculated for the plane electron wave in the usual way and they exhibit the ZB
oscillations (see Ref. \cite{Zawadzki06}). For small momenta $k_y$ the ZB frequency
is $\hbar\omega_Z=E_g$, where $E_g=2\hbar uk_{n\nu}$.
The ZB amplitude is $\lambda_Z\approx 1/k_{n\nu}$.
However, we are again interested in the displacement $\bar{y}(t)$ of a charge carrier
represented by a one-dimensional wave packet analogous to that described in Equation (\ref{defPacket}),
see Ref. \cite{Rusin07b}.
Figure \ref{Fig5} shows the ZB oscillations calculated for a Gaussian wave packet of two widths.
The decay times are of the order of pikoseconds, i.e. much larger
that in bilayer and monolayer graphene. The reason is that
the ZB oscillations occur due to the 'built in' momentum $k_x$ arising from the tube's topology.
In other words, the long decay time is due to the one-dimensionality of the system.

The last subject we consider is the trembling motion of electrons in monolayer graphene
in the presence of an external magnetic field $\bi{B} \parallel z$, see \cite{Rusin07c}.
The magnetic field is known to cause no interband electron transitions,
so the essential features of ZB, which results from an
interference of positive and negative energy states of the system, are expected not to be destroyed.
On the other hand, introduction of an external field provides an important parameter affecting the ZB behavior.
The essential feature introduced by the magnetic field is a quantization of the electron spectrum
$E_{ns}=s\hbar\omega\sqrt{n}$, where $n=0,1,\ldots$ and $s=\pm 1$ for
the conduction and valence bands, respectively.
The basic energy is $\hbar\omega=\sqrt{2}\hbar u/L$, with $1/L=(eB/\hbar)^{1/2}$.
The velocity operators can again be calculated in the time-dependent Heisenberg picture. These are
subsequently averaged over a Gaussian wave packet and they exhibit the trembling motion.
The presence of a quantizing magnetic field has very important effects on ZB:
(1) For $B\neq 0$ the ZB oscillations are {\it permanent}, for $B=0$ they are transient.
(2) For $B\neq 0$ many ZB frequencies appear, for $B=0$ only one frequency is at work. (3) For $B\neq 0$
both interband and intraband (cyclotron) frequencies contribute to ZB, for $B=0$ there are no
intraband frequencies. (4) Magnetic field intensity changes not only the ZB frequencies but the entire
character of ZB spectrum.

\section{Discussion}
It follows from the recent theoretical papers that the phenomenon of ZB should appear quite
often in solids.
Whenever one deals with interacting energy bands, charge carriers having a non-vanishing momentum should
experience the trembling motion. Its frequency $\omega_Z$ is about $\hbar\omega_Z \approx \Delta {\cal E}$,
where $\Delta {\cal E}$ is the energy separation between the interacting 'positive' and 'negative' states,
and the amplitude $\lambda_Z$ goes as $\lambda_Z \propto 1/\Delta {\cal E}$. The latter can reach the
values of tens of nanometers in narrow-gap semiconductors.
As mentioned in the Introduction, a similar phenomenon should occur for free relativistic electrons
in a vacuum, but both its frequency and amplitude are much less favorable than for electrons in
semiconductors, see Refs. \cite{Thaller04,ZawadzkiHM}.
This correspondence of the trembling motions illustrates
a general analogy between the behavior of electrons in narrow-gap semiconductors and that of relativistic
electrons in a vacuum \cite{ZawadzkiHM}. The Zitterbewegung was also proposed for other systems
\cite{Clark08,Zhang08PRL}, and it has some interesting analogies in quantum optics \cite{Bermudez08}.
As to the ways of observing the ZB, one can try
to detect directly the moving charge at corresponding frequencies using the scanning probe microscopy
\cite{LeRoy03,Topinka00}. One can also try to measure the ac current related
to the oscillating charge, see Figures \ref{Fig3}, \ref{Fig4} and Ref. \cite{Rusin07c}).
Finally, and this is probably the most promising way,
the oscillating charges should produce an observable dipole radiation. An external magnetic field
can continuously change the frequency of such radiation.
It appears that graphene in
the presence of a magnetic field provides most favorable conditions for an
observation of the fascinating phenomenon of Zitterbewegung.
Very recently, an acoustic analogue of Zitterbewegung was  observed
in a two-dimensional sonic crystal \cite{Zhang08}.

\ack
One of us (W.Z.) acknowledges many years of friendship and collaboration with Guenther Bauer.
We are both pleased and honored to be able to contribute to the {\it Festschrift } celebrating
the 65th birthday of Professor Bauer.

\end{document}